\documentclass[aps,pre,twocolumn,showpacs,showkeys,groupedaddress]{revtex4-1}

\usepackage{amsmath}
\usepackage{amssymb}

\usepackage{graphicx}
\usepackage{acronym}

\usepackage{soul} 
\usepackage{color}

\newcommand{\spindir}{{\boldsymbol\sigma}}
\newcommand{\RR}{{\boldsymbol R}}

\begin{document}

\title{Ageing and crystallization in a lattice glass model}



\author{Alejandro Seif}

\email{seifalejandro@inifta.unlp.edu.ar}

\author{Ernesto S. Loscar}

\author{Tom\'as S.~Grigera}

\thanks{Present address: Instituto de F\'\i{}sica de L\'\i{}quidos y
  Sistemas Biol\'ogicos (IFLYSIB --- CCT La Plata --- CONICET), Calle
  59 no.\ 789, B1900BTE La Plata, Argentina}

\affiliation{Instituto de Investigaciones Fisicoquımicas Te\'oricas y
  Aplicadas (INIFTA) and Departamento de F\'\i{}sica, Facultad de
  Ciencias Exactas, Universidad Nacional de La Plata, c.c.~16, suc.~4,
  B1904DPI La Plata, Argentina}

\affiliation{CCT CONICET La Plata, Consejo Nacional de Investigaciones
  Cient\'\i{}ficas y T\'ecnicas, Argentina}

\date{December 8, 2014}

\begin{abstract}

  We have studied a the 3-$d$ lattice glass of Pica Ciamarra, Tarzia,
  de Candia and Coniglio [Phys.\ Rev.\ E. \textbf{67,} 057105 (2013)],
  which has been shown to reproduce several features of the structural
  glass phenomenology, such as the cage effect, exponential increase
  of relaxation times and ageing.  We show, using short-time dynamics,
  that the metastability limit is above the estimated Kauzmann
  temperature.  We also find that in the region where the metastable
  liquid exists the aging exponent is lower that 0.5, indicating that
  equilibrium is reached relatively quickly.  We conclude that the
  usefulness of this model to study the deeply supercooled regime is
  rather limited.

\end{abstract}


\maketitle

\acrodef{KMC}{Kinetic Monte Carlo}
\acrodef{TTI}{time-traslation invariance}

\section{Introduction}

\label{sec:introduction}

The physics of structural glasses and glass-forming liquids
\cite{review:cavagna09}, in particular fragile liquids
\cite{review:angell95}, is still an open problem
\cite{review:biroli13}.  Several theoretical explanations have been
put forward to explain the sharp slowdown that supercooled liquids
experience near the glass transition temperature
\cite{review:tarjus05, review:cavagna09,
  review:berthier2011,review:chandler10}, as well as other concomitant
dynamic and thermodynamic features, but no single one has gained
widespread acceptance.  Part of the problem is that distinguishing
among theories requires data very difficult to obtain from experiment.

Numerical simulations have been heavily used to investigate this
problem \cite{rev:kob_leshouches2002}, employing models ranging from
realistic to minimal.  A minimal model should exhibit the basic
phenomenology of glasses while allowing simplified theoretical study
and/or fast numerical simulation (slow dynamics being usually an
obstacle for numerical studies and preventing thermalization at
temperatures where the most important observations would have to be
made).  Lattice models belong naturally in the last category, and
several have been studied so far.

Here we consider again a lattice model proposed a few years ago: the
monodisperse lattice glass introduced by Pica Ciamarra, Tarzia, de
Candia and Coniglio \cite{lg:ciamarra03, lattice-glass:ciamarra03}
(henceforth PCTCC).  This model is attractive theoretically because it
is amenable to approximate study under the Bethe lattice scheme, and
numerically because it can be studied with \ac{KMC} without making
approximations, so that simulations can be carried to very long times
($10^{12}$ Monte Carlo steps or more).  This model has been shown to
reproduce the cage effect and slow dynamics (described by Mode
Coupling Theory \cite{rev:goetze92, review:das04}) in appropriate
density ranges, with power-law diffusion coefficient and stretched
exponential decay of time correlations \cite{lg:ciamarra03,
  lattice-glass:ciamarra03}), as well as dynamical heterogeneity
\cite{lg:decandia10, lg:kerrebroeck06}.  It also exhibts a random
first order transition \cite{mosaic:kirkpatrick89} on the Bethe
lattice \cite{lg:biroli01, lg:ciamarra03} (\textsl{i.e.}\ a Kauzmann transition
with vanishing of complexity, like the $p$-spin
model \cite{mosaic:kirkpatrick87b, p-spin:crisanti92}).

We reexamine the phenomenology of this model with an emphasis on deep
supercooling and aging behavior.  Since this model, as the real
materials it tries to emulate, has a stable crystal phase, the
(metastable) liquid cannot be found at arbitrarily low temperatures.
Not only does the metastable phase eventually loose stability (at the
thermodynamic spinodal \cite{review:binder04, spinodal:binder07}), but
in finite dimension it ceases to be observable (\textsl{i.e.\ } the
\emph{metastability limit}\/ is reached) \emph{before} it becomes
unstable \cite{review:kauzmann48}, at a point called
\emph{pseudospinodal,} or \emph{kinetic spinodal}
\cite{phase-transition-theory:patashinskii79,
  phase-transition-theory:patashinskii80, self:jcp03a}.  This is
defined as the point where the relaxation time of the liquid equals
the time it takes for a stable crystal nucleus to form.  Since the
liquid relaxation time is growing rapidly in these systems, the
location of the kinetic spinodal arguably deserves more attention than
it is usually granted.  This should be especially the case in lattice
models, where one does not expect the elastic effects that may, in
real liquids, depress the kinetic spinodal enough that the liquid be
well defined down to the Kauzmann temperature
\cite{nucleation:cavagna05}.  This issue is also relevant for
out-of-equilibrium (aging) studies.  The scaling exponent $\nu$ has
been shown to depend on how far from equilibrium the system actually
is \cite{ageing:warren12}, and at too high temperatures the asymptotic
regime $\nu=1$ may never be reached.  On the other hand, aging at
temperatures below the kinetic spinodal (too low temperatures) has a
completely different phenomenology, namely that of coarsening
\cite{self:jcp03a}.

We accordingly seek to establish under which values of the control
parameters the supercooled liquid is well defined in this model.  We
also reconsider here its aging behavior, which has up to here received
less attention.

The paper is organized as follows: In Sect.~\ref{sec:introduction} we
introduce the model and some details about our method of simultation,
in Sect.~\ref{sec:model-simulations} we present the study of the limit
of stability and aging, and in Sect.~\ref{sec:conclusions} we
conclude.


\section{Model and simulations}

\label{sec:model-simulations}

The PCTCC was introduced by Pica Ciamarra, Tarzia, de Candia and
Coniglio in refs.~\onlinecite{lg:ciamarra03,
  lattice-glass:ciamarra03}. It can be formulated as follows:
classical particles with an orientation (``spin'') are placed on a
simple cubic lattice of side $L$ where the occupation number of site 
$\RR$ is called $n_\RR$. In each site the orientation
$\spindir_\RR$ is a unit vector that can point in the direction of one
of the six first neighbors. Hard excluded-volume constraints
are imposed such that a) only one particle can occupy a given site
($n_\RR=0,1$) and b) the orientation vector must point to an empty
site ($\RR+\spindir=\RR'$ only if $n_{\RR'}=0$). 
In the canonical ensemble, the hard potential means that temperature 
does not play a role, and the control parameter of the model is the 
density $\rho=N/L^3$ ($N=\sum_\RR n_\RR$). 
On the other hand, in the grand canonical ensemble, the control parameter 
is the dimensionless Lagrange multiplier $\alpha=\beta\mu$ (with
$\beta$ the inverse temperature and $\mu$ the chemical potential).  
In this paper we will use the grand canonical ensemble and thus 
in this case, to draw parallels with normal micromolecular liquid, it is
useful to plot in terms of temperature; fixing $\mu=1$
and vary $\alpha$ through $\beta=1/T$. For the sake of uniformity, 
we'll be describing the results obtained with this ensemble in terms 
of $T$ or $1/T$.

The PCTCC has a known crystal state \cite{lattice-glass:ciamarra03}
for a cubic lattice with periodic boundary conditions.  It can be
built with the following rule: for each site $\RR=(x,y,z)$ evaluate
$a = (x + 2y + 3z) \mod 7$, then
\begin{itemize}
\item if $a=0$  leave site empty,
\item if $a=1,2,3$ place a particle pointing in negative $x$, $y$ or
  $z$ direction respectively,
\item if $a=4,5,6$ place a particle pointing in positive $x$, $y$ or z
  direction respectively.
\end{itemize}
The crystal has a density of $\rho = 6/7 \simeq 0,86$ (specific
volume $v=\rho ^{-1} \simeq 1.167$) and the unit cell is
$7\times7\times7$ sites.

To quantify the amount of crystal phase present in a given sample
define the \emph{crystal mass fraction} $m$ as the fraction of empty
sites surrounded by six particles pointing towards them (which is the 
only way that empty sites appear in the perfect cristal).
This quantity is very easy to evaluate and gives a measure of the 
amount of crystal, independent of the size of domains. It is not a 
proper order paramenter, since it will be nonzero also in the liquid 
phase, but as we shall see it increases significantly as the system 
starts crystallizing and it is a useful measure to detect the onset of
crystallization.

To study the dynamics we will consider the self-overlap $Q(t,t_w)$,
defined by
\begin{equation}
  \label{eq:3}
  Q(t,t_w)= \sum_\RR n_\RR(t+t_w) n_\RR(t_w) \spindir_\RR(t+t_w)\cdot  \spindir_\RR(t_w).
\end{equation}
$Q(t,t_w)$ is a measure of the memory of the configuration at time $t_w$
retained at time $t_w+t$. It is independent of $t_w$ if the system is in
equilibrium.

\subsection{Simulations}

\begin{figure}
  \includegraphics[width=\columnwidth]{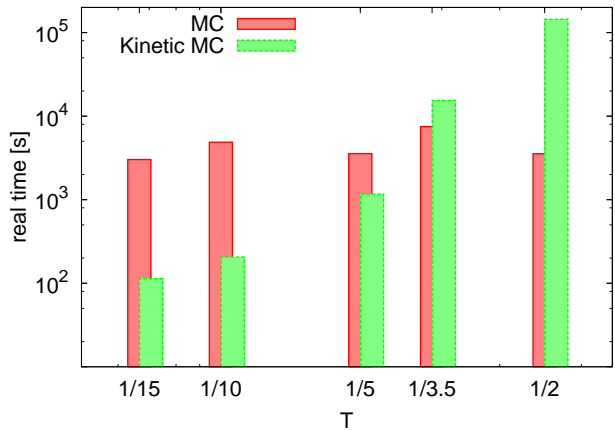}
  \caption{Comparison of real simulation times of standard Monte Carlo and
    kinetic Monte Carlo.  Kinetic MC is highly dependent on the
    density, which is directly related to the available actions that the
    system can perform. \label{Benchmarking}}
\end{figure}

At high densities (which is the regime of interest), the model
evolves very slowly because the number of allowed moves is very small.
Under these conditions, standard Metropolis Monte Carlo is very
inefficient, since most of the moves proposed are ultimately rejected.
Thus we resort to Kinetic Monte Carlo (KMC) \cite{algorithm:bortz75}
(also known as ``the $n$-fold way'', dynamic Monte Carlo, or Gillespie
algorithm \cite{algorithm:gillespie77}).  The idea is to compute the
probability of a transition out of the current configuration (which
will be very small at high densities) and force a move to one of the
possible destination configurations, advancing the time by the inverse
of the total transition probability.  The actual transition performed
is selected at random from a list of all possible moves.  Although at
the beginning of a simulation with an empty lattice the list of moves
is very long, once the system starts filling up with particles the
move list becomes smaller and smaller, thus speeding the simulation.
Clearly, using KMC for low density systems is a bad idea, since the
additional bookkeeping required to maintain a long list is more
time-consuming than a simple Metropolis Monte Carlo. The algorithm
consists of the following steps:
\begin{enumerate}
\item Compile a list of the $M$ possible moves and their probabilities
  $p_i$.
\item \label{item:1} Perform a move randomly selected from the list (weighted by its
  probability $p_i$).
\item Advance time by $1/\sum_i p_i$.
\item Update the list of moves and go to step \ref{item:1}.
\end{enumerate}
For our system with hard constraints $\sum_i p_i \sim 1/M$, so that
for small $M$ each step advances time by a large amount. Standard
Monte Carlo (MC) performs at approximately the same speed per MC time
unit independent of $T$, as it is shown in Fig.~\ref{Benchmarking}.
In contrast, KMC algorithm with $L=14$ and $T > 1/5$ (\textsl{i.e.}\
in the fast liquid regime) is 10 to 100 times slower than MC. However,
below $T = 1/5 = 0.2 $ the time decreases exponentially and KMC is 1
or 2 orders of magnitude faster than MC. Since we are studying the
whereabouts of the glass transition, which we study for $T < T_m
\simeq 0.15$ (see below), we are working in the region where KMC is
faster than MC.

\section{Results}

\begin{figure}
  \includegraphics[width=\columnwidth]{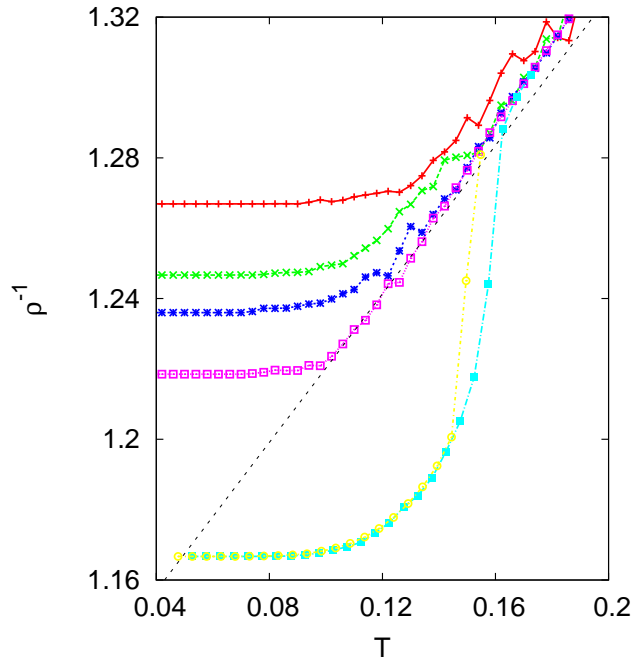}
  \caption{$\rho^{-1}$ as a function of $T$ for different cooling
    rates. For $\dot{T}=-10^{-5}$ (red) the system
    goes quickly out of equilibrium, while $\dot{T}=-10^{-6}$ (green),
    $\dot{T}=-10^{-7}$ (blue) and $\dot{T}=-10^{-8}$ (purple) allow the
    supercooled metaequilibrium liquid to be found at progressively lower temperatures.
    Heating from the perfect crystall can be used to estimate $T_m$,
    as in the cyan ($\dot{T}=10^{-8}$) and yellow ($\dot{T}=10^{-9}$)
    curves.}
\label{HeatingCooling}
\end{figure}

As shown in ref.~\onlinecite{lg:ciamarra03}, this model is very slow
to crystallize.  If one prepares the system in the pure crystal state,
one can estimate the melting temperature $T_m$ through slow heating.
The data shown in Fig.~\ref{HeatingCooling} for this slow heating
gives $T_m \simeq 0.15$.  In contrast, upon cooling no sign of
crystallization is seen and the system remains in a supercooled liquid
state until it goes out of equilibrium at a cooling-rate-dependent
temperature (Fig.~\ref{HeatingCooling}).  One can extrapolate the
specific volume curve of the supercooled liquid branch (dashed line in
Fig.~\ref{HeatingCooling}) and find its intersection with the crystal
value. The corresponding temperature, $T_K\simeq 0.05$, can be used as
an estimate of the Kauzmann temperature \cite{review:kauzmann48}.

\begin{figure}
  \includegraphics[scale=0.7]{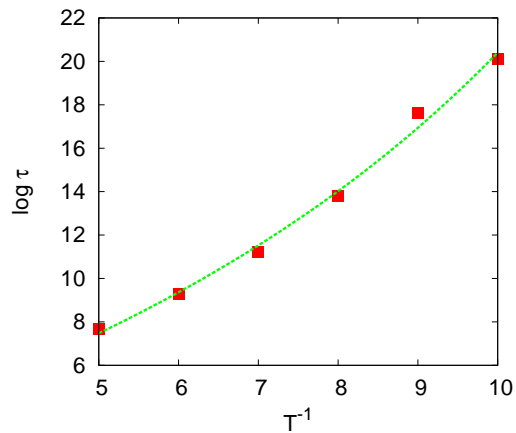}%
  \caption{Angell plot showing non-Arrhenius behavior.  Full curve is
    a Vogel-Fulchnet-Tamman fit with $T_0=0.04$.}
 \label{TauAlpha}
\end{figure}

Relaxation times $\tau_R$ can be obtained by means of the self overlap
$Q(t,t_w)$.  These times, ploted in Fig.~\ref{TauAlpha}, are well
fitted with a Vogel-Fulchner-Tamman law
\begin{equation}
\tau_R(T)=~\tau_0~\exp [B/(T-T_0)],
\end{equation}
with $T_0 =0.04$, quite close to $T_K$.

\subsection{Metastability limit}

\begin{figure}
  \includegraphics[scale=0.7]{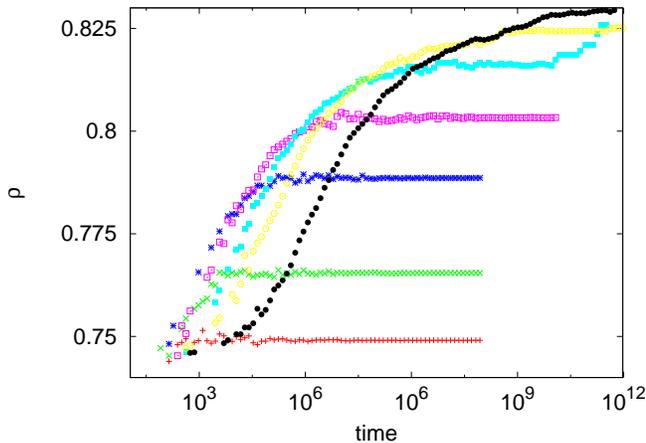}%
  \caption{Density vs.\ time for (from bottom to top) $T=1/5$,
  1/6, 1/7, 1/8, 1/9, 1/10, and 1/12.}
  \label{Density}
\end{figure}

\begin{figure}
  \includegraphics[scale=0.7]{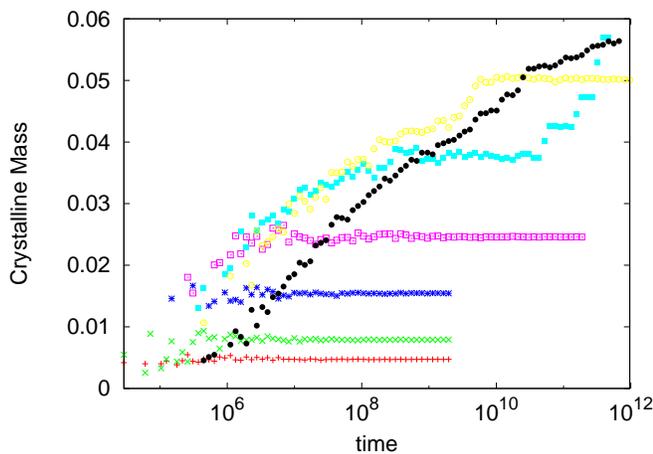}%
  \caption{Crystal mass fraction for (from bottom to top) $T=1/5$,
  1/6, 1/7, 1/8, 1/9, 1/10, and 1/12.}
  \label{CrystalMass}
\end{figure}

We now attempt to establish the lowest temperature at which the
supercooled liquid can be equilibrated, \textsl{i.e.\ }the
metastability limit. We have done quenches from the empty lattice to
several values of $T_f$ (Fig.~\ref{Density}). For $T_f \ge 1/8$, the
lattice fills up relatively quickly until the density reaches a
$T$-dependent plateau after $\sim 10^3$--$10^5$ steps. The crystal
mass fraction $m$ increases more slowly but also reaches a plateau
(Fig.~\ref{CrystalMass}). The plateau regime is candidate (subject to
aging checks, see sec.~\ref{sec:aging}) for the equilibrium
(meta-equilibrium for $T<T_m\approx 0.15$) liquid.  This being a
short-range model, however, the meta-equilibrium state cannot be
expected to last forever, and indeed at $T=1/9$ one clearly sees that
$\rho$ and $m$ leave the plateau after about $10^{10}$ steps and
continue increasing towards the crystal values.  We interpret this as
a crystal growth regime, where one or more supercritical crystal
nucleii have formed and are slowly growing. The system is no longer
liquid, but out of equilibrium again.

For $T_f=1/12$, however, the behavior is different: the growth of
$\rho$ and $m$ is slower, but a plateau is never reached.  Instead,
both quantities continue to grow towards the crystal values, reaching
relatively high values more quickly than systems at higher values of
$T_f$. The system is never in a metastable state, instead entering a
coarsening regime before the metastable liquid can equilibrate.  We
can thus take $1/12 \simeq 0.083$ as a lower bound for
$T_{\text{sp}}$.  The estimated Kauzmann point, at $T_K=0.05$ is thus
way past the metastability limit, making it of questionable relevance.

\subsubsection{Short-time dynamics}

To locate the thermodynamic spinodal, which serves as a lower bound on
the metastability limit, we used the short-time dynamics technique as
recently proposed \cite{self:jcp09}.  The technique is based on the
fact that the thermodynamic spinodal is an instability similar to a
critical point, but located in the metastable region.  Then this
instability can be exploited \cite{self:jcp09} to locate the spinodal
studying the critical short-time dynamics \cite{short-time:janssen89,
  review:zheng06, self:rpp11}. The procedure consists in looking for a
power-law time relaxation from an initial state prepared according to
some prescription. In equilibrium critical points the power-law regime
lasts for a time increasing with the system size, but in the case of
spinodals (in a sense metastable critical points) this regime is found
only for a finite interval \cite{self:jcp09}.  The procedure was as
follows:
\begin{enumerate}
\item Prepare a well equilibrated sample at high temperature
  $T_{\textsl{i}}=4T_{\textsl{c}}$. This is the disordered initial state.
\item At $t=0$ quench suddenly to $T_{\textsl{f}}\lesssim
  T_{\textsl{c}}$.  Let the system relax while recording the order
  parameter and its fluctuations up to $t\sim 10^6\,$MCS (\emph{short
    time}).
\item Look for power-law behavior.  The spinodal temperature is
  determined as that where the power-law regime lasts the longest.
\end{enumerate}
We chose the crystall mass fraction $m(t)$ as order parameter and
along with it computed the sample-to-sample fluctuations
\begin{equation}
  \chi_m(t) = N \sigma_m(t) = N \sqrt{\langle [m(t)- \langle m(t) \rangle ]^{2}\rangle} , 
\end{equation}
where $\langle\ldots\rangle$ stands for average over thermal history
and sample (starting configuration). The normalized fluctuation
$\chi_m$ should be independent of the system size and invariant under
a shift of $m$, avoiding the problem that $m$ is not a proper order
parameter for the spinodal point.  Thus, in this point we expect a
pseudocritical dynamics given by $\chi_m(t) \propto t^{\phi}$.

\begin{figure}
\begin{center}
\includegraphics[width=\columnwidth,clip]{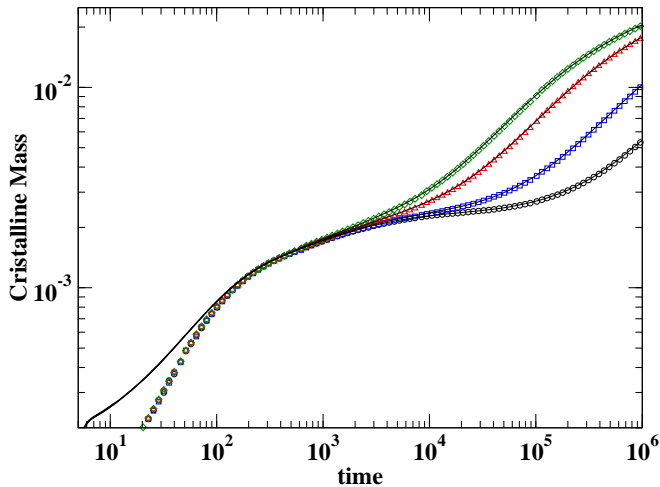}
\caption{Evolution of the crystal mass fraction starting from the
  disordered initial condition ($T_{\textsl{i}}=4T_{\textsl{c}}$) with
  final temperatures $T_f=0.08$ (circle), $0.09$ (square), $0.104$
  (triangle), and $0.112$ (diamond), and using a system of side
  $L=21$. Continuous lines are the same temperatures for a bigger
  system of side $L=30$.}
\label{fig:mass}
\end{center}
\end{figure}

Fig.~\ref{fig:mass} shows the time evolution of the mass fraction $m$,
starting at configurations prepared at
$T_{\textsl{i}}=4T_{\textsl{c}}$, for different temperatures (in the
supercooled region) and using systems of side $L=21$, and $L=30$. This
quantity always increases, being a good parameter to detect the onset
of the process of forming the solid phase. Note that size effects
disappear for $t>10^2$, and also the technique starts to distinguish
different temperatures for $t>10^3$.

\begin{figure}
  \includegraphics[width=\columnwidth,clip]{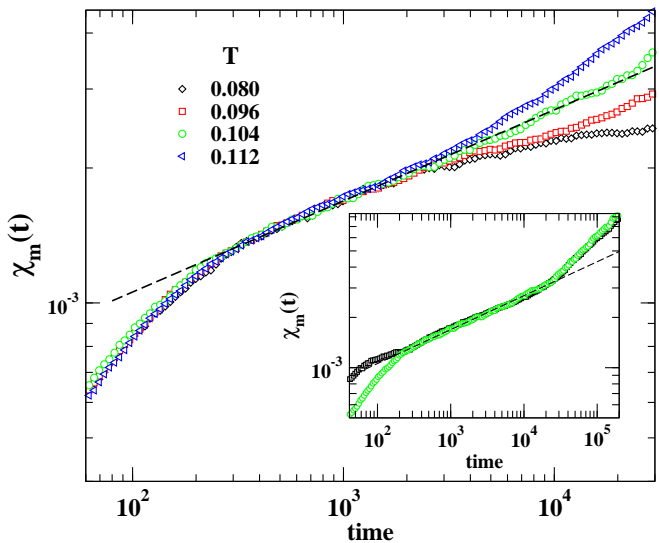}
  \caption{Evolution of the normalized fluctuations for different
    temperatures (as indicated) for a system of side $L=30$. For
    $T=T_{sp}=0.104$ we obtain a power law behavior, the dashed black
    line is a fit of a power law with exponent $\phi=0.203$. The inset
    shows results for $T=T_{sp}$ and two system sides $L=21$ (black
    squares) and $L=30$ (green circles). }
  \label{fig:sigmamass}
\end{figure}

Fig.~\ref{fig:sigmamass} shows $\chi_m(t)$ vs.\ $t$ for a system of
side $L=30$ at different temperatures. These data are obtained with
$10^4$ runs. We can see a power law behavior in the range $200\leq t
\leq 27000$ (more than two decades) at $T=T_{sp}=0.104$. The power law
fit gives an exponent $\phi=0.20 \pm 0.01$. In the inset of
Fig.~\ref{fig:sigmamass} we plot a comparison of $\chi_m(t)$ for a
system of $L=21$, showing again that the results are independent of
$L$ for $t>200$. From this data we estimate the temperature for the
thermodynamic spinodal point as
\begin{equation}
T_{sp}= 0.104\pm0.004.
\end{equation}
This temperature is about twice $T_K$, confirming our earlier
statement that the Kauzmann point is irrelevant in this system.

%
%
\subsection{Aging}

\label{sec:aging}

To study aging, we consider the self overlap $Q(t,t_w)$ as a function
of two times for waiting times $t_w\ge10^6$, starting with the lattice 
empty, in the region of slow approach to the plateau of the density 
(Fig.~\ref{Density}).

\begin{figure}
  \includegraphics[width=\columnwidth]{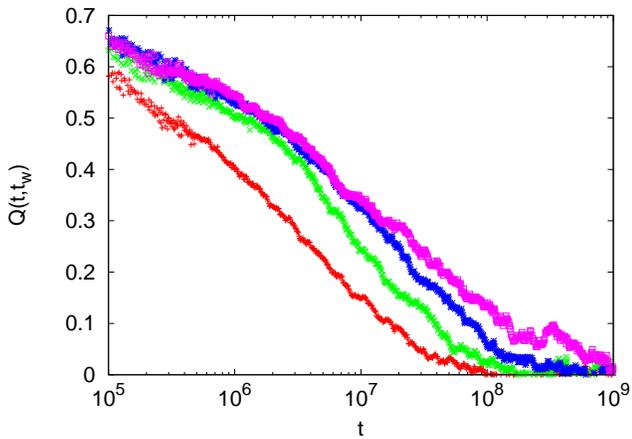}
  \caption{Overlap vs.\ time at $T=1/9$ for $t_w=10^6$ (red),
    $t_w=10^7$ (green) $t_w=2 \times 10^8$ (blue) and $t_w=2 \times
    10^9$ (purple).}
  \label{PCTCCAgeing}
\end{figure}

For $T>1/8$ we find that \ac{TTI} holds, i.e.\ $Q(t,t_w)\equiv
Q(t-t_w)$ and consequently no ageing is observed (not shown).  On the
other hand, for $1/T=9$, we find $t_w$ dependence
(Fig.~\ref{PCTCCAgeing}).  The system shows signs of aging from
$t_w=10^6$ to $t_w\sim 10^8$; after that the curves start becoming
close to each other.  This interruption of ageing coincides roughly
with the appearence of the plateau in $\rho$ and $m$
(Figs.~\ref{Density} and~\ref{CrystalMass}), and is an indication that
the liquid is equilibrating.  This plateau lasts up to $t\sim
10^{11}$, when the system leaves equilibrium again to begin
crystallizing.

\begin{figure}
  \includegraphics[width=\columnwidth]{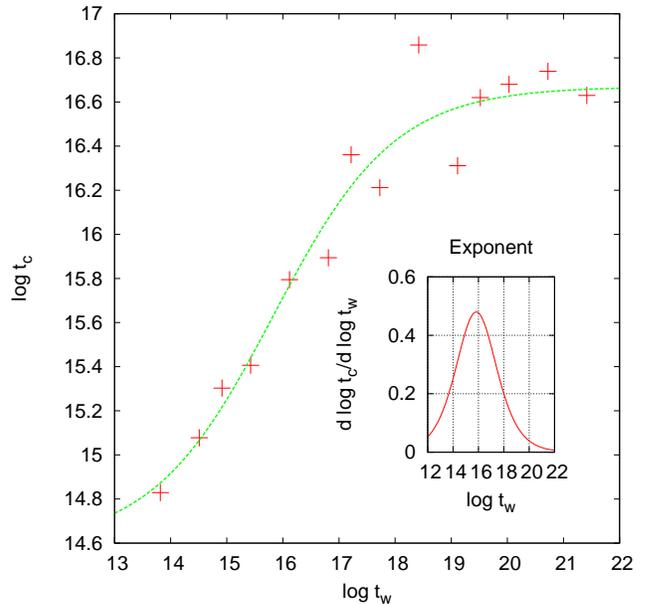}
  \caption{Characteristic time of the relaxation ($t_c$) vs.\ $t_w$
    for $T=1/9$.  The full line is a fit to a 4 parameter sigmoid
   $S(t)=A+\frac{B}{1+e^{-(t/D-C)}}$.  Inset: effective exponent
    $\nu_\text{eff} = \partial \ln t_c / \partial \ln t_w$.}
  \label{PCTCCScaling}
\end{figure}

In the aging regime, structural glasses have been found to obey the scaling
\begin{equation}
  \label{eq:2}
  Q(t,t_w)=f\left(\frac{t}{t_w ^{\nu}}\right),
\end{equation}
with $\nu$ close to 1 \cite{ageing:struik77, self:prb04}.  This
relation clearly cannot apply to the data of Fig.~\ref{PCTCCAgeing}
across all waiting times, it is because the curves coincide for
$t>3\times10^9$.  We can however compute an effective scaling exponent
$\nu(t_w)$ \cite{ageing:struik77, ageing:warren12}:  Defining a
characteristic decay time $t_c$ using a fixed thershold for the
overlap (we chose $Q(t_c) = 0.3$), one can define an effective scaling
exponent through $\nu_\text{eff}(t_w) =
\partial \ln t_c / \partial \ln t_w$.  Rather than evaluating the
derivative numerically, we use a sigmoidal fit for $t_c$ vs $t_w$
\cite{ageing:warren12}; see Fig.~\ref{PCTCCScaling}.  The low
values of $\nu$ for $t_w=10^8$ confirm that the liquid is reaching
(meta)equilibrium.  However, $\nu$ is never higher than 0.5, quite far
from the value $\nu\approx1$. According to
ref.~\cite{ageing:warren12}, and taking also into account the behavior
of the density, we interpret this as evidence that the model is always
``too close'' to equilibrium to reproduce the experimental aging
behavior.  In other words, the dynamics is starting to become slow, so
that it takes a relatively long time for the system to equilibrate,
but there is no proper aging regime; equilibrium is reached relatively
soon after one-time quantities begin to stabilize (while
experimentally aging in the correlation function happens with very
little variation of one-time quantities \cite{ageing:struik77}.  The
situation is similar at the other temperatures we have studied (above
and near the metastability limit), see
Table~\ref{nu-alpha-scaling-PCTCC}.

\begin{table}
\begin{tabular}{ l | r }
	\hline 
  $T$ & $\nu_\text{max}$\\
  \hline     
1/8 & 0.441 \\
1/9 & 0.476 \\
1/10 & 0.455 \\
\hline  
\end{tabular}
\caption{Estimate of maximum value of the aging exponent $\nu$ for
  different temperatures}
\label{nu-alpha-scaling-PCTCC}
\end{table}

Of course, at lower temperatures the dynamics will be slower, and it will
take longer to reach equilibrium.  However, as shown above
temperatures beyond $0.10$ are below the metastability limit, so that
the out-of-equilibrium behavior in that region corresponds to a
coarsening regime, where aging is qualitatively different from a
structural glass \cite{review:corberi11}.

\section{Conclusions}

\label{sec:conclusions}

We have revisited the Pica Ciamarra-Tarzia-de Candia-Coniglio lattice
glass.  While it reproduces many features of supercooled liquids, as
previously pointed out, the present analysis shows that it is not
suitable to study the deeply supercooled regime.  We have shown that
the metastability limit is at a temperature not too far from the
melting point, making the metastable liquid nonexistant for
$T\lesssim 0.10$, above the estimated Kauzmann temperature.  Although
thermodynamic studies, and in particular theories relating dynamic
behavior to thermodynamic properties are not invalidated by an
unreachable Kauzmann point (as long as the metastable liquid exists),
in the present model the range of validity of such studies seems to be
too restricted.

Studies of the structural glass (out of equilibrium) state are also
somewhat limited.  Beyond the metastability limit, the out of
equilibrium behavior is that of coarsening, in principle rather
different than what has been observed in experimental and numerical
studies of structural glasses.  The range between $T=0.10$ and melting
indeed corresponds to the structural glass situation, i.e. an
out-of-equilibrium system slowly evolving towards an equilibrium
liquid.  However, we have found that the $t/t_w$ scaling is not obeyed
by this model, probably because it reaches equilibrium too quickly in
this temperature range.  When cooled further, relaxation becomes slower
and one would hope to get something closer to a structural glass;
unfortunately the liquid ceases to exist before a regime with the
correct scaling sets in.


\begin{acknowledgments}
We thank R.~A.~Borzi, D.~A.~Martin and G.~Parisi for useful suggestions and discussions.
\end{acknowledgments}

\bibliographystyle{apsrev4-1}
\bibliography{rabib}

\end{document}